%% file: ms.tex
\documentclass[twocolumn]{aastex631}
\usepackage{showyourwork}
\usepackage[ruled,vlined,linesnumbered]{algorithm2e}
\usepackage{algorithmic}
\usepackage{color}
\usepackage{booktabs}
\usepackage{xspace}

\definecolor{rb4}{HTML}{27408B}

\definecolor{pyRed}{RGB}{214, 39, 40}

\newcommand{\flatiron}{\affiliation{Center for Computational Astrophysics,
Flatiron Institute, New York, NY 10010, USA}}
\newcommand{\jhu}{\affiliation{William H. Miller III Department of Physics and Astronomy, Johns Hopkins University, Baltimore, Maryland 21218, USA}}

\SetKwInput{Parameters}{Parameters}
\SetKwInput{Variables}{Variables}

\begin{document}

\title{Fast gravitational wave parameter estimation without compromises}

\author{Kaze W. K. Wong}
\email{kwong@flatironinstitute.org}
\flatiron

\author{Maximiliano Isi}
\flatiron

\author{Thomas D. P. Edwards}
\jhu

\begin{abstract}
We present a lightweight, flexible, and high-performance framework for inferring
the properties of gravitational-wave events. By combining likelihood
heterodyning, automatically-differentiable and accelerator-compatible waveforms,
and gradient-based Markov chain Monte Carlo (MCMC) sampling enhanced by
normalizing flows, we achieve full Bayesian parameter estimation for real events
like GW150914 and GW170817 within a minute of sampling time.  Our framework does
not require pre-training or explicit reparameterizations and can be generalized
to handle higher dimensional problems. We present the details of our
implementation and discuss trade-offs and future developments in the context of
other proposed strategies for real-time parameter estimation. Our code for running the analysis is publicly available on
GitHub \url{https://github.com/kazewong/jim}.
\end{abstract}

\newcommand{\ptot}{}
\newcommand{\jsdMaxBBH}{}
\newcommand{\jsdAvgBBH}{}
\newcommand{\jsdMaxBNS}{}
\newcommand{\jsdAvgBNS}{}
  \input{output/macros.tex}\label{output/macros.tex}\unskip%

\section{Introduction}
\label{sec:intro}

Parameter estimation (PE) underpins all of gravitational-wave physics and
astrophysics, and is one of the most commonly performed tasks in
gravitational-wave (GW) data analysis \citep{Christensen:2022bxb,
2019PASA...36...10T}. The central goal of PE is to infer the parameters of a
particular GW source given the strain data recorded by instruments like LIGO
\citep{LIGOScientific:2014pky,LIGOScientific:2021usb,LIGOScientific:2021djp}, Virgo \citep{VIRGO:2014yos} and KAGRA
\citep{KAGRA:2020tym}.  In the standard compact binary coalescence (CBC)
scenario, this could mean inferring intrinsic parameters such as the masses and
spins of the compact objects, as well as extrinsic parameters such as their sky
localization and distance from Earth. PE is also applied to test general
relativity (GR)
\citep{LIGOScientific:2016lio,LIGOScientific:2018dkp,LIGOScientific:2021sio},
and constrain the properties of nuclear matter \citep{LIGOScientific:2018hze}.
PE is a crucial step in GW science, since it translates characteristics of the
strain data into astrophysically relevant quantities that can be used to
constrain astrophysical phenomena, including informing theories of binary
evolution \citep{LIGOScientific:2021psn}.

There exist a number of prominent, community-developed PE codes, including
\textsc{LALInference} \citep{Veitch:2014wba}, \textsc{PyCBC Inference}
\citep{Biwer:2018osg}, and \textsc{Bilby}
\citep{Ashton:2018jfp,Romero-Shaw:2020owr}.  These packages have been tested by a
number of groups and are well regarded as standard tools. However, while these
tools have passed many robustness tests, they are known to be computationally
intensive. The exact amount of time needed to analyze one event depends on
factors like the duration and frequency of the signal, as well as features of
the specific waveform model. Typical runtimes for production-level analyses can
range from hours to weeks. This expense precludes iterating quickly on results,
launching large scale measurement simulations, or obtaining results in low
latency to inform astronomers for potential followup in real time.

Additionally, in the coming decade, there are planned upgrades for existing
facilities, as well as plans for next-generation detectors such as the Einstein
Telescope (ET) \citep{Punturo:2010zz} and the Cosmic Explorer (CE)
\citep{LIGOScientific:2016wof}. These upgrades will increase the instrument's sensitivity
and allow for the detection of more events with a better
signal-to-noise ratio (SNR). The number of events that will be detected in the
coming decade is expected to grow from around a thousand per year to over a
million per year \citep{Baibhav:2019gxm}. This will put a significant strain on
the current PE tools.

In order to address this, there are efforts from multiple groups to speed up the
PE process. This includes methods that employ techniques such as
reduced-order quadrature \citep{Canizares:2014fya,Smith:2016qas}, adaptive proposal distributions in nested sampling
\citep{Williams:2021qyt}, deep learning networks pre-trained on large collections
of waveforms \citep{Dax:2021tsq,Dax:2022pxd}, as well as methods that reduce the
computational expense of classical PE by leveraging our knowledge of GW signals
\citep{Veitch:2014wba,Ashton:2021anp,Cornish:2021wxy,Islam:2022afg,Roulet:2022kot,Lee:2022jpn,Lange:2018pyp,Wofford:2022ykb}.
While these avenues are promising for standard GW problems, they rely on
assumptions that may not hold for analyses targeting additional physical effects
beyond standard CBCs in GR, such as matter effects, lensing and deviations from GR.


In this work, we present a lightweight, flexible, and robust
framework to infer GW event parameters in a fully-Bayesian analysis. Our
approach relies on the following techniques to achieve its performance:
\begin{enumerate}
\setlength{\itemsep}{0pt}
\item likelihood heterodyning,
\item differentiable waveform models,
\item a normalizing-flow enhanced Markov chain Monte-Carlo (MCMC) sampler, and
\item native support for hardware accelerators.
\end{enumerate}
All these components, working jointly, come together into a high-performance
and high-fidelity PE pipeline, which can achieve ${\gtrsim}10^3\times$ speed
ups relative to traditional tools without compromising accuracy or generality
for efficiency, and without making limiting assumptions about the target
parameter space.

The rest of the paper is structured as follows: we review the basics of PE and
introduce our framework in Sec.~\ref{sec: PE}; we present benchmarking results
on both simulated and real data in Sec.~\ref{sec: Result}; and, finally, we
discuss the implications of this work and directions for future development in
Sec.~\ref{sec: Discussion}.

\section{Gravitational wave parameter estimation}
\label{sec: PE}

\subsection{Likelihood function}
\label{sec:likelihood}

The main objective of PE is to obtain a multidimensional posterior distribution
$p(\mathbf{\theta} \mid d)$ on parameters $\mathbf{\theta}$ given strain
data $d$.  This probability density represents our best inference of
the source properties, and encodes all relevant information contained in the
observed data.
To compute this object, we use Bayes' theorem to write
\begin{align} \label{eq: bayes}
    p(\theta \mid d) = \frac{\mathcal{L}(d \mid \theta)\pi(\theta)}{p(d)}\, ,
\end{align}
where $\mathcal{L}(d \mid \theta)$ is the likelihood function, $\pi(\theta)$ is
the prior distribution, and $p(d)$ is the evidence. Since the evidence is a
normalization constant that does not depend on the source parameters, it is
often omitted if we are only interested in the posterior distribution. The prior
distribution is often chosen to be something simple (e.g., uniform in the component masses or a Gaussian distribution in the spins), or
it could directly encode astrophysical information. Assuming the noise is drawn from a
Gaussian process, the log-likelihood for GW data is given by
\begin{align}
    \log{\mathcal{L}(d \mid \theta)} = -\frac{1}{2} \left\langle d-h(\mathbf{\theta}) \mid d-h(\mathbf{\theta})\right\rangle,
\label{eq: loglikelihood}
\end{align}
where $d$ is the observed strain data, and $h(\mathbf{\theta})$ is the signal
predicted by a waveform model with a specific set of source parameters
$\theta$.  The right hand side of Eq.~\eqref{eq: loglikelihood} can be
evaluated in either the time or frequency domains. For stationary noise, it is
computationally cheaper to compute the likelihood in the frequency domain, with
a noise-weighted inner product given by
\begin{align}
    \left\langle a \mid b\right\rangle = 4 \Re \int_0^\infty \frac{a^*(f)b(f)}{\mathcal{S}_n(f)}\, \mathrm{d}f \, ,
\label{eq: innerproduct}
\end{align}
where $\mathcal{S}_n(f)$ is the one-sided power spectral density (PSD) of the
noise.  In practice, the integral becomes a discrete sum over a finite number
of bins determined by the sampling rate of the detector data and duration of
the observation.

To compute the integral in Eq.~\eqref{eq: innerproduct}, we need to evaluate a
waveform model $h(\mathbf{\theta})$ at a number of frequency bins. This makes
evaluating the likelihood function often the most computationally intensive
part of PE. The most accurate waveforms are obtained via numerical relativity
(NR) simulations \citep{BaumgarteShaprio:NumRel}, which directly solve the Einstein equations numerically for a
given system. Unfortunately, such simulations can take anywhere from a day to
half a year, making direct evaluation through NR prohibitively expensive. To
circumvent this, there are several families of waveform ``approximants'',
including the \textsc{IMRPhenom} family \citep{Khan:2015jqa,
Garcia-Quiros:2020qpx}, the \textsc{SEOB} family \citep{PhysRevD.89.061502}, and
the NR surrogate family \citep{Varma:2019csw}. Since PE requires millions of
likelihood evaluations during sampling, the computational cost in evaluating
the waveform is a major contributor to the long runtimes of GW PE.

\subsection{Heterodyned likelihood}

Since the cost of evaluating a waveform model scales linearly with the number
of time or frequency bins, the computational burden for longer-duration signals
is often quite large. To mitigate this, there are a number of methods to reduce
the number of basis points needed to compute the likelihood faithfully
\citep{Field:2011mf, Field:2013cfa, Smith:2016qas,
Vinciguerra:2017ngf,Morisaki:2020oqk,Morisaki:2021ngj}. In this work, we take
advantage of likelihood heterodyning \citep{Cornish:2010kf,Cornish:2021lje}
(also known as relative binning \citep{Zackay:2018qdy}).

The idea behind the heterodyned likelihood can be summarized as follows: since
the integrand in Eq.~\eqref{eq: innerproduct} is a highly oscillatory function,
one has to sample it at a high rate to compute the integral faithfully;
however, the number of sample points needed would be much smaller if the
integrand was smooth. For a pair of points $\theta$ and $\theta_0$ that are
close to each other in parameter space, the corresponding waveforms $h(\theta)$
and $h(\theta_0)$ will necessarily be similar; this means that the ratio
between waveforms is a smooth function of frequency. Given a reference
waveform $h(\theta_0)$, we can exploit this fact to reduce the number of
frequency bins needed to compute the likelihood for the set of $\theta$ in the
neighborhood of $\theta_0$.

To do this, we decompose the integrand into two parts: (1) a highly oscillatory
part that depends only on the reference waveform given by $\theta_0$ and
the data, and which need only be evaluated once; and, (2) a smoothly
varying part that depends on the target waveform parameters $\theta$, which
must be evaluated for every likelihood computation. Because the part
that depends on the target waveform parameters is smooth, we can use far fewer
frequency samples to compute the integral with sufficient accuracy.

One may be concerned about the accuracy of this scheme, especially in the
region where the generated waveform is significantly different from the
reference waveform. However, given that we are interested in the most probable
set of parameters, if we choose the reference waveform to be close to the data,
the waveforms that are different from the reference waveform will necessarily
also differ significantly from the data. This means that the likelihood value
for parameters far from the reference will be significantly smaller than the
likelihood of those close to it, and hence will not be relevant for the PE
result. To ensure that this is the case, we always pick reference parameters
known to lie close to the target, e.g., by first maximizing the likelihood
function using the highest frequency resolution available, which can be run at
a much lower cost than full PE. 

We now give a concise description of the implementation of this approach in our
code; for a more extensive derivation of heterodyned likelihood, we refer the
reader to \cite{Zackay:2018qdy}. Let $h(f)$ and $h_0(f)$
represent the target and reference waveforms, respectively; then, for a given
sparse binning of the frequency axis, the ratio $r(f) = h(f)/h_0(f)$ can be
well approximated by a linear interpolation over the bin,
\begin{align}
r(f) \approx r_0(h,b) + r_1(h,b)(f- f_m(b)) + \cdots,
\label{eq:definer}
\end{align}
where $b$ is the index of a particular bin, $r_0(h,b)$ and $r_1(h,b)$ are
respectively the value and slope of the ratio at the center of the bin, and
$f_m(b)$ is the central frequency of the bin. Since we have access to both
$h(f)$ and $h_0(f)$, we can compute $r_0$ and $r_1$ by evaluating $r(f)$ at the
edge of the bin and inverting Eq.~\eqref{eq:definer}.

With this definition, the two terms involving $h$ obtained by expanding
Eq.~\eqref{eq: loglikelihood} can be approximated as
\begin{subequations} \label{eq: heterodynedlikelihood}
\begin{equation}
    \langle d \mid h \rangle \approx \sum_b \left[ A_0(b)\, r^*_0(h,b) + A_1(b)\, r^*_1(h,b) \right] ,
\end{equation}
\begin{align}
    \langle h \mid h \rangle \approx \sum_b &\left[ B_0(b)\, |r_0(h,b)|^2 + \right. \nonumber \\
    &\left. 2 B_1(b)\, \Re\{r_0(h,b)\, r_1(h,b)\} \right] 
\end{align}
\end{subequations} 
where $A_0(b)$, $A_1(b)$, $B_0(b)$, and $B_1(b)$ are heterodyning coefficients
computed using the data and the reference waveform. These are defined to be
\begin{subequations} \label{eq:hetcoeff}
\begin{align}
    A_0(b) &= 4 \sum_{f \in b} \frac{d(f)h^*_0(f)}{S_n(f)} \Delta f, \\
    A_1(b) &= 4 \sum_{f \in b} \frac{d(f)h^*_0(f)(f-f_m(b))}{S_n(f)} \Delta f, \\
    B_0(b) &= 4 \sum_{f \in b} \frac{|h_0(f)|^2}{S_n(f)} \Delta f, \\
    B_1(b) &= 4 \sum_{f \in b} \frac{|h_0(f)|^2(f-f_m(b))}{S_n(f)} \Delta f\, ,
\end{align}
\end{subequations}
where the sums within each bin ($f \in b$) should be done with the same, dense
sampling rate as the original data (with thin frequency bins of width $\Delta f$).

To evaluate Eq.~\eqref{eq: heterodynedlikelihood}, we need to first choose a
binning scheme, then evaluate the coefficients in Eq.~\eqref{eq:hetcoeff} given
the data and the reference waveform, and at last the ratio between the target
waveform and the reference waveform at the center of each bin via
Eq.~\eqref{eq:definer}.

The phasing of an inspiral waveform is denoted by a power series
$\Psi(f) = \sum_i \alpha_i f^{\gamma_i}$, where $\alpha_i$ are some
coefficients depending on the waveform parameters and $\gamma_i$ are powers
motivated by post-Newtonian theory. For example, for the term $\gamma_i =
-5/3$, $\alpha_i$ is related to the chirp mass. The maximum dephasing one can
have within a frequency interval $[f_{\textrm{min}},f_{\textrm{max}}]$ is 
\begin{align}
    \delta \Psi_{\textrm{max}}(f) = 2\pi \sum_{i} (f/f_{*,i})^{\gamma_i} \textrm{sgn}(\gamma_i),
\label{eq: maxdephasing}
\end{align}
where $f_{*,i} = f_{\textrm{max}}$ for $\gamma_i \geq 0$ and $f_{*,i} =
f_{\textrm{min}}$ for $\gamma_i<0$. Given the relation shown in Eq.~\eqref{eq:
maxdephasing}, we can choose the binning scheme to divide the entire frequency
band of interest into a set of bins such that the maximum dephasing within each
bin is smaller than a certain threshold $\epsilon$, i.e.,
$|\delta\Psi_{\textrm{max}}(f_{\textrm{max}}) -
\delta\Psi_{\textrm{max}}(f_{\textrm{min}})| < \epsilon$. 


To obtain a reference waveform, we currently use the \textsc{differential
evolution} algorithm \citep{Storn1997DifferentialE} available in the
\textsc{scipy} package \citep{2020SciPy-NMeth} to find the waveform parameters
which maximize the likelihood.  The reference waveform could also be produced
from trigger parameters precomputed by a search pipeline without additional
computation. Once we have obtained a reference waveform, we can check the accuracy
of the heterodyned likelihood by comparing its value to the original likelihood
at several points in the parameter space. We can then choose the number of bins such
that the difference between the values of the two likelihoods is smaller than
chosen tolerance threshold.

\subsection{MCMC with Gradient-based sampler}
\label{sec:gradient}

Given Eq.~\eqref{eq: loglikelihood} and the prior, one can evaluate the
posterior density, Eq.~\eqref{eq: bayes}, over the entire parameter space of
interest to obtain the most probable set of values that are consistent with the
data. However, directly sampling this posterior quickly becomes intractable as
the dimensionality of the parameter space increases beyond a few dimensions.
Markov chain Monte Carlo \citep{gelmanbda04} is a common method employed to
generate samples from the target posterior when direct sampling is not
possible.

In MCMC, the posterior distribution is approximated by a Markov chain that
eventually converges to the target distribution \citep{10.1214/aos/1176325750}.
The chain is constructed by iteratively proposing a new point in the
parameter space based on the current location of the chain. The proposed point
is accepted with a probability that is usually set to be proportional to the
ratio of the posterior density evaluated at the proposed point and the current
point. The chain can either accept the proposal and move to the new location,
or reject the proposal and stay at the current location. This process is
repeated until the chain converges to the target distribution. The samples
generated by the chain are then used as a fair sample to estimate the
quantities of interest, such as the mean and credible intervals of the source
parameters. In practice, since we do not know the target distribution ahead of
time, the MCMC process is usually repeated until a certain criterion is met,
such as a Gelman-Rubin convergence statistic \citep{10.2307/2246093} lower than
certain threshold, or simply after a fixed number of iterations.

Compared to direct sampling, MCMC algorithms only explore regions that are
highly probable, thus reducing the computational cost by not wasting resources
on parameters that are unlikely to generate the observed data. However, MCMC
algorithms come with their own set of issues. To illustrate what difficulties
MCMC may face, we can examine one of the most standard MCMC algorithms: the
Metropolis-Hastings algorithm with a Gaussian kernel. Starting at some initial
point, one can draw a proposed point from a Gaussian transition kernel, defined
as
\begin{align}
    q(\mathbf{x},\mathbf{x_0})= \mathcal{N}(\mathbf{x} \mid \mathbf{x_0},\mathbf{C}),
\end{align}
where $\mathbf{x_0}$ is the current location of the chain, $\mathbf{x}$ is the
proposed location, and $\mathbf{C}$ is the covariance matrix of the Gaussian. In
the simplest case, we can pick $\mathbf{C}$ to be a diagonal matrix with a
constant value, which corresponds to an isotropic Gaussian center around the
current location and with a fixed variance. The acceptance criterion is
defined as
\begin{align}
\alpha(\mathbf{x},\mathbf{x_0}) = \min\left(1,\frac{p(\mathbf{x})q(\mathbf{x_0},\mathbf{x})}{p(\mathbf{x_0})q(\mathbf{x},\mathbf{x_0})}\right).
\label{eq:Gaussian_acceptance}
\end{align}

We can see from Eq.~\eqref{eq:Gaussian_acceptance} that the acceptance rate is
proportional to the fraction of volume where the posterior density at the
proposed location is higher than the current location within the Gaussian
transition kernel. If we choose the variance of the transition kernel to be too
large, this fraction will be small hence the acceptance rate will be poor. On
the other hand, if one chooses the variance to be too small, nearby samples
will be correlated, and it will take a long time for the chain to wander. In
both cases, the efficiency in constructing a chain with a target number of
independent samples is suboptimal. Consequently, there is often a tuning
process before we run the MCMC algorithm to find the optimal sampling settings
(in this example, the variance of the Gaussian) to ensure the best possible
performance. 

However, as we often deal with high-dimensional problems, even the optimally
tuned Gaussian transition kernel does not guarantee good performance. In order
to have a reasonable acceptance rate, the variance of the Gaussian has to be
smaller in a higher dimensional space, which means that the transition kernel
will generally make smaller and smaller steps as we increase the dimensionality
of the problem \citep{2017arXiv170102434B}.

Transition kernels that leverage gradient information of the target
distribution can help address this issue of shortening steps in a high
dimensional space. Instead of proposing a new point by drawing from a Gaussian,
one can use the gradient evaluated at the current location to propose a new
point, so that the evolution of the chain is preferentially directed to regions
of higher probability. For example, the Metropolis-adjusted Langevin algorithm
(MALA) \citep{10.2307/2346184} places a unit Gaussian at the tip of the gradient
vector at the current position,
\begin{align}
    \mathbf{x} = \mathbf{x_0} + \tau \nabla\log{p(\mathbf{x_0})} + \sqrt{2\tau}N(0,\mathbf{I})\, ,
\end{align}
where $\tau$ is a step size chosen during the tuning stage. Compared to a
Gaussian centered at the current location, the MALA transition kernel is more
likely to propose a point in the higher posterior density region because of the
gradient term, which helps boost the acceptance rate.
Hamiltonian Monte Carlo \citep{2017arXiv170102434B} is another gradient-based
algorithm that has been explored for neutron star inspirals
\citep{Bouffanais:2018hoz}.

While transition kernels that use gradient information can help improve the
acceptance rate, computing the gradient of the posterior density function
introduces an additional computational cost, which is not necessarily
beneficial in terms of sampling time. If one wants to compute the gradient
through finite differencing, the additional computational cost goes as at least
${\sim} \mathcal{O}(2n)$, where $n$ is the dimension of the problem. On the
other hand, schemes like \textsc{Jax}
\citep{jax2018github} allow us to compute the gradient of the likelihood
function with respect to the parameters through automatic differentiation,
which gives the gradient information down to machine precision at around the
same order of time compared to evaluating the posterior itself. Thus, having
access to gradient information through automatic differentiation is crucial to
making gradient-based transition kernels favorable in terms of computing cost.

To leverage automatic differentiation, the entire posterior function must be
implemented within \textsc{Jax} or a similar framework; this includes the
likelihood and, therefore, the waveform approximant.  This means that the
development of differentiable approximants is essential for leveraging
gradient-based sampling in GW applications. We currently make use of the
waveforms implemented in \textsc{ripple} \citep{ripplepaper}.

\subsection{Normalizing Flow enhanced sampling}
\label{sec:flow}

While gradient-based samplers have been shown to outperform gradient-free
algorithms in many practical applications, there remain classes of problems
that most gradient-based samplers do not solve well. For example, first-order
gradient-based algorithms struggle with target distributions that exhibit
locally-varying correlations, since they assume a single mass matrix that does
not depend on the location of the chain by construction
\citep{2017arXiv170102434B}.\footnote{Sampling algorithms that use higher order
derivatives such as manifold-MALA and Riemannian-HMC \citep{RMHMC} can in
principle handle local correlations in the target distribution; however, they
often encounter instabilities when used in real-life applications, so their use
is a rare practice.} Another example is multi-modality: if there are multiple
modes in the target distribution, individual chains will likely be trapped in
one mode and take an extremely long time to transverse between the modes
\citep{2018arXiv180803230M}.  This means that the relative weights between modes
will take much longer to estimate than the shape of each mode.

Moreover, before we can use the sampling chain to estimate the posterior
quantities we care about, the sampler often needs to first find the most
probable region in the target space (known as the \emph{typical set}); this is
a common process often referred to as ``burn-in'' in the literature. As a
consequence, one would discard a certain amount of data generated from the
beginning of the sampling process, and only use the later part of the chain to
estimate the quantities of interest. The burn-in phase of a gradient-based
sampler is often as long as the sampling phase, which means that a good portion
of the computation is not directly devoted to estimating the target quantities.

All the above issues can be mitigated by normalizing flows.  Normalizing flows
are a technique based on neural networks that aims at learning a mapping from a
simple distribution, such as a Gaussian, to a complex distribution, often given
in the form of samples \citep{2019arXiv190809257K, 2019arXiv191202762P}. Once
the network is trained, one can evaluate the probability density of the complex
distribution and sample from it very efficiently, by first evaluating the
simple distribution and then applying the learned mapping. The core equation
of normalizing flows is the coordinate transformation of probability
distributions via a Jacobian, as given by
\begin{align}
    p_x(X) = p_z(Z) \left| \frac{\partial f}{\partial z}\right|^{-1},
\end{align}
where $p_x(X)$ is the complex target distribution, $p_z(Z)$ is the simple
latent distribution and $f$ is an invertible parameterized transform that
connects the two distributions, $x = f(z)$, to be learned by the normalizing
flow. See \cite{2019arXiv190809257K, 2019arXiv191202762P} for a detailed
discussion of the algorithm.

Working in tandem, gradient-based MCMC and normalizing flows can efficiently
explore posteriors with local and global correlations, as well as multiple
separate modes.  The scheme relies on iteratively using draws from the
gradient-based MCMC to train a normalizing flow, which is then itself used as a
proposal for another stage of MCMC sampling.

Concretely, we begin by producing initial training data for the normalizing
flow by running multiple independent chains of the gradient-based algorithm for
a fixed number of steps.  From the resulting pool of samples, the normalizing
flow can begin to learn the landscape of the target distribution.
However, since the independent chains contain the same number of samples, the
relative weight assigned to each chain will not represent the true target
distribution (e.g., the relative importance of separate modes will not be
correctly calibrated). This is mitigated by a second stage of gradient-based
MCMC sampling that uses the distribution learned by the normalizing flow as a
\textit{proposal}.

Given a trained normalizing flow model, we can generate the proposed jump in
the target space by sampling from the latent distribution $z \sim p_z(Z)$,
usually a Gaussian, and then pushing it through the learned map given by the
normalizing flow model $x=f(z)$.  The acceptance criterion is then set to be
\begin{align} \label{eq:flow-proposal}
    \alpha(\mathbf{x},\mathbf{x_0}) = \min \left[ 1, \frac{\hat{\rho}(\mathbf{x_0})\rho_*(\mathbf{x})}{\hat{\rho}(\mathbf{x})\rho_*(\mathbf{x_0})}\right],
\end{align}
where $\hat{\rho}$ is the probability density estimated by the normalizing flow
model, $\rho_*$ is the probability density evaluated using the target function,
and $x_0$ is the current position.

From Eq.~\eqref{eq:flow-proposal}, we can see that the flow distribution is the
target distribution when the accepting probability is 1. When the normalizing
flow model has not converged to the target distribution, only a portion of the
proposed jumps will be accepted. This means an MCMC process using the
normalizing flow model as the proposal distribution can adjust the
normalization across different regions of the target parameter space by
rejecting jumps into less likely regions. The training and sampling are then
repeated until certain criteria are met, at each step combining global and
local MCMC sampling which respectively do and do not use the normalizing flow
as proposal.

Note that every time we retrain the network, we are breaking the Markov
properties since we are changing the proposal distribution. To produce final
samples that can be used to estimate target quantities, one has to freeze the
normalizing flow model and not retrain during the final sampling phase in order
to satisfy the detailed balance condition. We use the package \textsc{flowMC}
\citep{2022arXiv221106397W,Gabrie:2021tlu}, with MALA as the gradient-based sampler. The
pseudocode of the algorithm is given in Algorithm \ref{alg:cap}.

\begin{algorithm}
\caption{\textsc{flowMC} pseudocode}\label{alg:cap}
\KwIn{initial position $ip$}
\Parameters{number of training loops $nt$, number of production loops $np$}
\Variables{current chains $cc$, current position $cp$, current NF parameters $\Theta$, chains from local sampler $c_{local}$, chains from global sampler $c_{global}$}
\KwResult{output chains $chains$}
$cp \leftarrow ip$\\
\tcc{Training loop}
    \For{$i<nt$}{
        $cc, cp \leftarrow LocalSampling(cp)$\\
        $\Theta \leftarrow TuneNF(cc)$\\
        $c_{global}, cp \leftarrow GlobalSampling(cp, \Theta)$ \\
        $cc \leftarrow Append(cc, c_{global})$
    }
\tcc{Production loop}
    \For{$i<np$}{
        $c_{local}, cp \leftarrow LocalSampling(cp)$\\
        $c_{global}, cp \leftarrow GlobalSampling(cp, \Theta)$ \\
        $chains \leftarrow Append(chains, c_{local}, c_{global})$
    }

\Return{$chains$}
\end{algorithm}

\subsection{Accelerators}
\label{sec:accelerators}

Modern hardware accelerators, such as graphics processing units (GPUs) and
tensor processing units (TPUs), are designed to execute large-scale, dense
computation. They are often much more cost-efficient than using many central
processing units (CPUs) when it comes to solving problems that can be benefited
from parallelization.  The downside of these accelerators compared to CPUs is
that they can only perform a more restricted set of operations and are often
less performant when dealing with serial problems. Parameter estimation with
MCMC is a serial problem since each new sample generated from a chain depends
on the last sample in the chain. This means that naively putting the problem on
an accelerator is more likely to reduce performance than increase it.

Yet, in our work, the use of accelerators provides two independent advantages
that tremendously benefit the parameter estimation process. First, using
accelerators allows us to run many independent MCMC chains simultaneously,
which benefits the training of the normalizing flow. Since we generate the data
we use to train the normalizing flow on the fly, the more independent data we
can feed to the training process, the higher chance the normalizing flow can
learn a reasonable representation of the global landscape of the target
distribution.  If we only used a small number of chains, we would be limited to
the correlated samples from each chain and we would have to run more sequential
steps to obtain the same amount of independent samples in the end---with more
chains the problem becomes parallelizable and we can obtain the same number of
training samples sooner. In other words, being able to use many independent
chains helps the normalizing flow learn the global landscape faster in wall
time.

Another benefit of accelerators is the parallel evaluation of waveforms. Since
the waveform model we use can be evaluated at any given time or frequency
independently, this means computing a waveform can be trivially parallelized
over frequency bins. On an \textsc{Nvidia} A100 GPU, we can evaluate the
waveform model $\sim \mathcal{O}(10^9)$ times in a second for different
frequencies or source parameters. Together with the heterodyned likelihood, this
allows us to run thousands of parallel chains in a PE run. The high throughput
of waveform evaluations unlocks the potential of the \textsc{flowMC} sampling
algorithm.

\section{Result}
\label{sec: Result}
\subsection{Injection-recovery test}

To demonstrate the robustness of our pipeline, we use it to recover the
parameters of a set of simulated signals injected into different instantiations
of synthetic stationary Gaussian noise. Then we run our pipeline on the
simulated data, and determine the credible interval at which the true
parameters of the injected signals are recovered. From the set of credible
values, we can check whether the truth lies within a certain credible interval
at the expected frequency: if our pipeline is working as expected, we should
find that the true parameters lie within $x\%$ credible interval $x\%$ of the
time, e.g., the true value should lie within the $50\%$ credible interval
$50\%$ of the time. In other words, the recovered percentiles of the true
parameters should be uniformly distributed. Deviation from this behavior would
suggest the pipeline is either over-confident or too conservative
\citep{Cook2006,Talts2018}.

We sample 1200 events from the distribution of parameters detailed in Table
\ref{tab:parameters}; the same distributions are used as the prior in the PE
process.  We simulate signals over $16\,$s of data, with a minimum frequency
cutoff of $30\,$Hz and a sampling rate of $2048\,$Hz. We draw noise from a set of
projected design PSDs for the LIGO Hanford, LIGO Livingston
(\texttt{SimNoisePSDaLIGOZeroDetHighPower}) and Virgo
(\texttt{SimNoisePSDAdvVirgo}) detectors
\citep{lalsuite,Shoemaker:T0900288,2012arXiv1202.4031M}. For both injection and
recovery, we make use of the \textsc{IMRPhenomD} waveform \citep{Khan:2015jqa}
via the fully-differentiable implementation presented in the \textsc{ripple}
package \citep{ripplepaper}. We use a neural spline flow model \citep{2019arXiv190604032D}
with 10 layers, each with 128 hidden units, and 8 bins per layer as our
normalizing flow model.

\begin{table*}[hbt!]
    \begin{center}
    \begin{tabular}{ l l l l l }
    \hline
    \hline
    Parameter &  Description & Injection & GW150914 & GW170817\\
    \hline

    $M_c$ & chirp mass $[M_\odot]$& $[10, 50]$ & $[10,80]$ & $[1.18,1.21]$ \\
    $q$ & mass ratio & $[0.5, 1]$ & $[0.125,1]$ & $[0.125,1]$ \\
    $\chi_1$ & primary dimensionless spin& $[-0.5, 0.5]$ & $[-1,1]$ & $[-0.05,0.05]$ \\
    $\chi_2$ & secondary dimensionless spin & $[-0.5, 0.5]$ & $[-1,1]$ & $[-0.05,0.05]$ \\
    $d_L$ & luminosity distance $[\textrm{Mpc}]$ & $[300, 2000]$ & $[0, 2000]^\dag$ & $[1, 75]^\dag$ \\
    $t_c$ & coalescence time $[\textrm{s}]$& $[-0.5, 0.5]$ & $[-0.1, 0.1]$ & $[-0.1, 0.1]$ \\
    $\phi_c$ & coalescence phase & $[0, 2\pi]$ & $[0, 2\pi]$ & $[0, 2\pi]$ \\
    $\cos{\iota}$ & cosine of inclination angle & $[-1, 1]$ & $[-1, 1]$ & $[-1, 1]$ \\
    $\psi$ & polarization angle & $[0, \pi]$ & $[0, \pi]$ & $[0, \pi]$ \\
    $\alpha$ & right ascension & $[0, 2\pi]$ & $[0, 2\pi]$ & $[0, 2\pi]$ \\
    $\sin{\delta}$ & sine of declination & $[-1, 1]$ & $[-1, 1]$ & $[-1, 1]$ \\

    \hline
    \hline
    \end{tabular}
    \caption{Prior ranges for parameters varied in the injection-recovery test,
    as well as the GW150914 and GW170817 analyses. All priors are uniform over
    the ranges shown, except for the luminosity distance prior in the GW150914
    and GW170817 analyses ($^\dag$) for which we apply a prior unform in
    comoving volume. The coalescence time refers to a shift relative to the
    geocenter trigger time, and $M_c$ refers to the redshifted (detector-frame)
    chirp mass.}
    \label{tab:parameters}
    \end{center}
\end{table*}

\begin{figure}
    \script{ppplots.py}
    \includegraphics[width=0.99\linewidth]{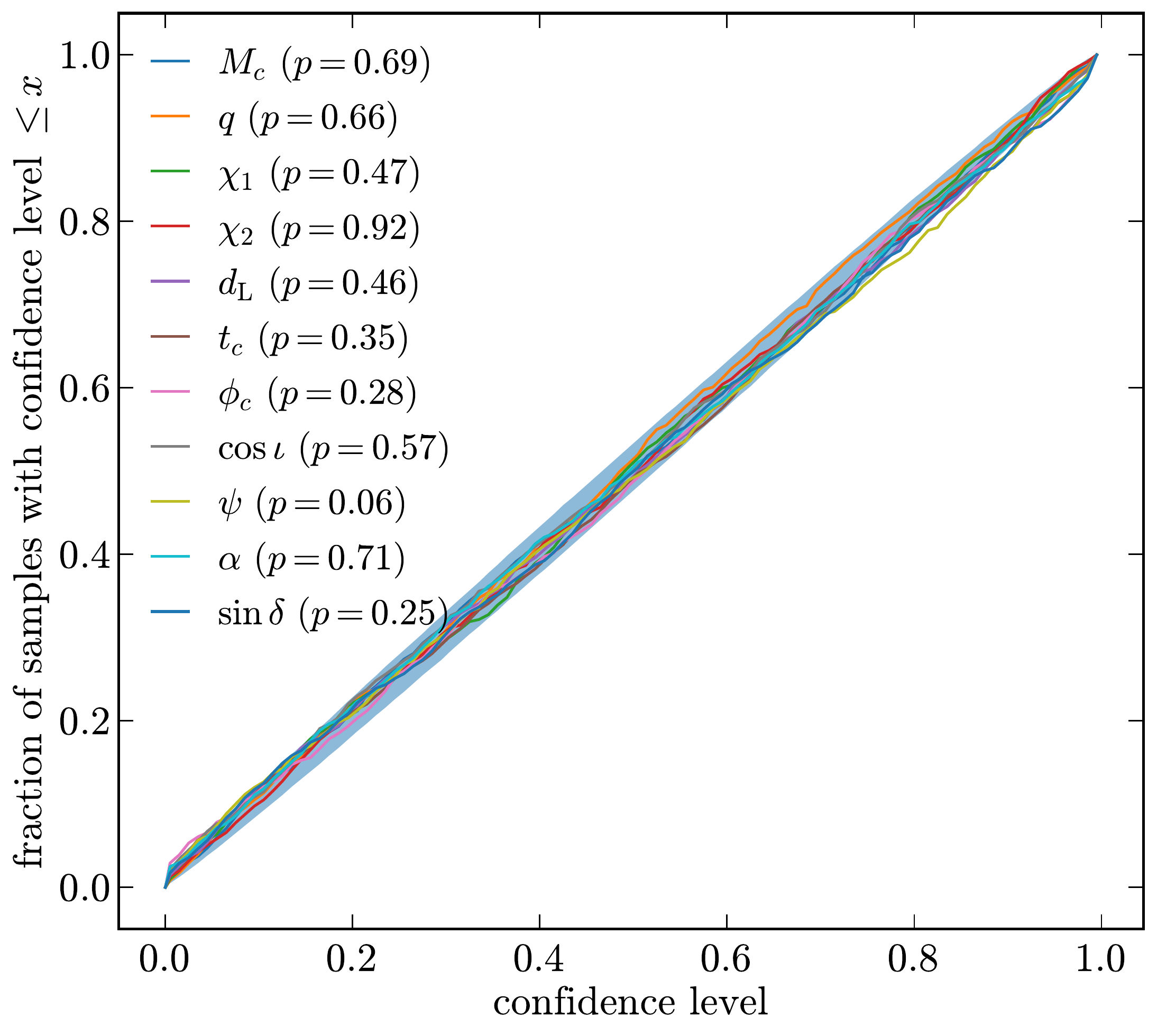}
    \caption{Cumulative distribution of the quantile of which the true value
    lies for each marginalized distribution. The shadow band denotes the 95\%
    credible interval drawn from a uniform distribution with the same number of
    events as the injection campaign. The legend shows the $p$-values for each
    marginalized distribution, with a combined value of $p = \ptot$.}
    \label{fig:ppplot}
\end{figure}

We summarize the result of this injection-recovery campaign in
Fig.~\ref{fig:ppplot}. This shows the cumulative distribution over injections
of the quantile at which the true value lies in the marginalized distribution
of each parameter. The shaded band denotes the 95\%-confident variation
expected from draws from a uniform distribution with the same number of events.
We can see that most of the measured curves lie within this band, showing that
our inference results agree well with a uniform distribution.

To further quantify how well our result agrees with a uniform distribution, we
can compute the Kolmogorov-Smirnov $p$-values for each marginalized
distribution \citep{doi:10.1080/01621459.1968.11009335}.
A low $p$-value (with a threshold often
chosen to be $p = 0.05$) could indicate that our result is in tension with a uniform
distribution. The $p$-values obtained for each parameter are shown in the
legend of Fig.~\ref{fig:ppplot}. Most of them are well above the $p = 0.05$
threshold, except for $\psi$, which is close to the threshold. Once
again, assuming these $p$-values are drawn from a uniform distribution, given
11 draws (the number of parameters in our inference), it is not abnormal to
have one of the parameters lying slightly outside the threshold. To assess
whether this is expected, we can compute the combined $p$-value for these 11
parameters, and find it to be $p = \ptot$.  This shows our inference pipeline
performs properly on simulated data at a similar level as standard tools
\citep{Veitch:2014wba,Romero-Shaw:2020owr}.

\subsection{Real event parameter estimation}

To demonstrate the performance of our parameter estimation pipeline, we apply
it to two real LIGO-Virgo events: GW150914 and GW170817. We use the priors
shown in Table \ref{tab:parameters}, and take 4 s of data sampled at 2048 Hz
starting at 20 Hz for GW150914, and 128 s of data sampled at 4096 Hz starting
at 23 Hz for GW170817; strain data and PSDs for both events are fetched from
GWOSC \citep{GWOSC}. We use the same normalizing flow model as the
injection-recovery study. For our specific choice of sampler settings, we
produce ${\sim}$2500 and 3500 \emph{effective samples}\footnote{Effective
samples here refers to the number of independent samples, which is the total
number of generated samples divided by their correlation length; we compute the
effective sample size using \textsc{arviz}
\citep{arviz_2019}, \url{https://python.arviz.org/en/stable/api/generated/arviz.ess.html}.
} for GW150914 and GW170817 respectively. Running on an \textsc{Nvidia} A100
GPU, the wall time for both events is around 3 minutes. Most of this time is
spent on just-in-time (JIT) compilation of the code; the actual sampling time
is only ${\sim}40$ s. We pre-compute the reference waveform parameters used to
hetetrodyne the likelihood for the two events, which is omitted in the
wall-time calculation.

\begin{figure*}
    \includegraphics[width=0.99\textwidth]{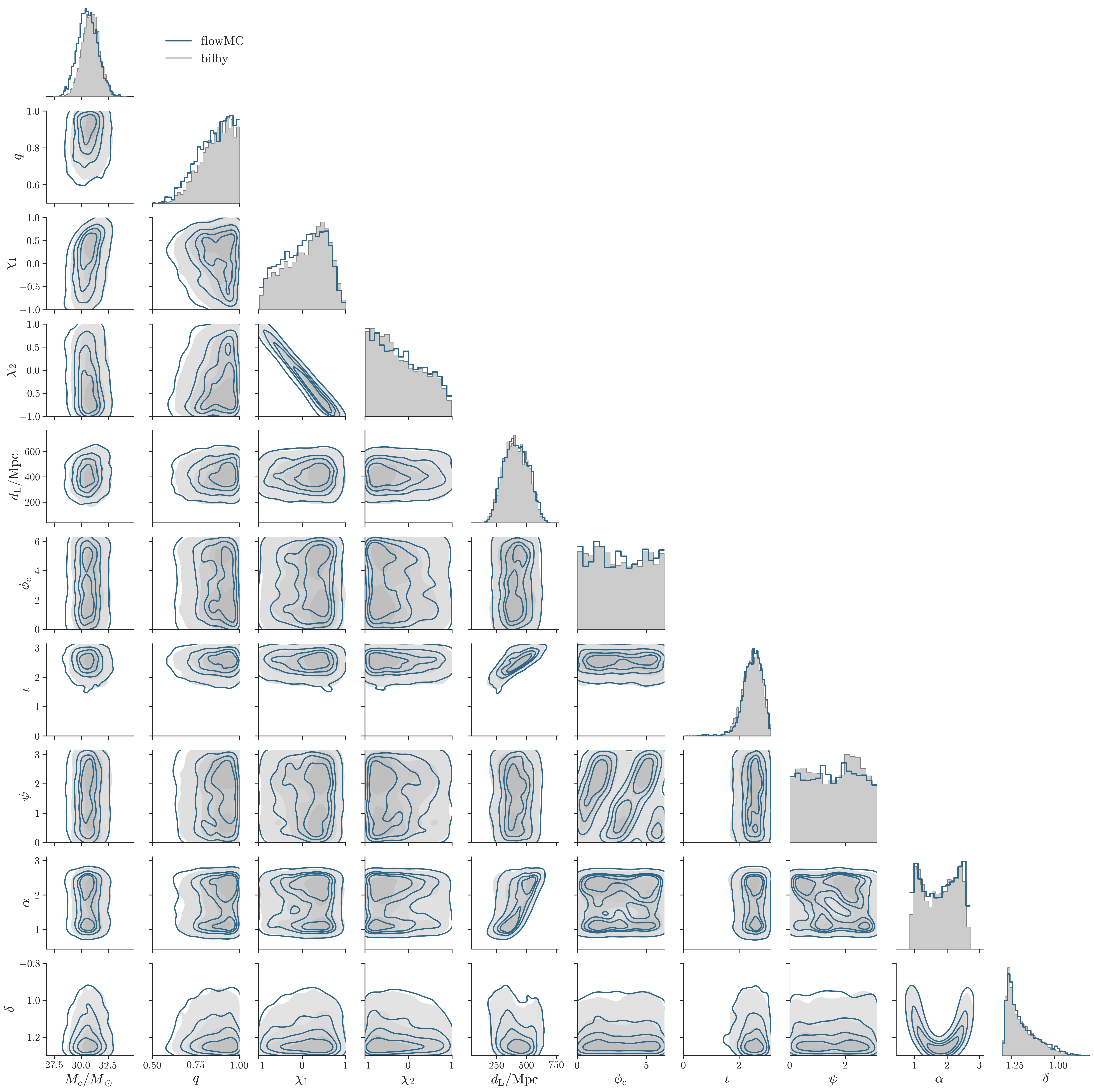}
    \caption{
      GW150914 posterior computed by our code (blue) and \textsc{Bilby} (gray).
    }
    \label{fig:GW150914}
    \script{plot_GW150914.py}
\end{figure*}

\begin{figure*}
\includegraphics[width=0.99\textwidth]{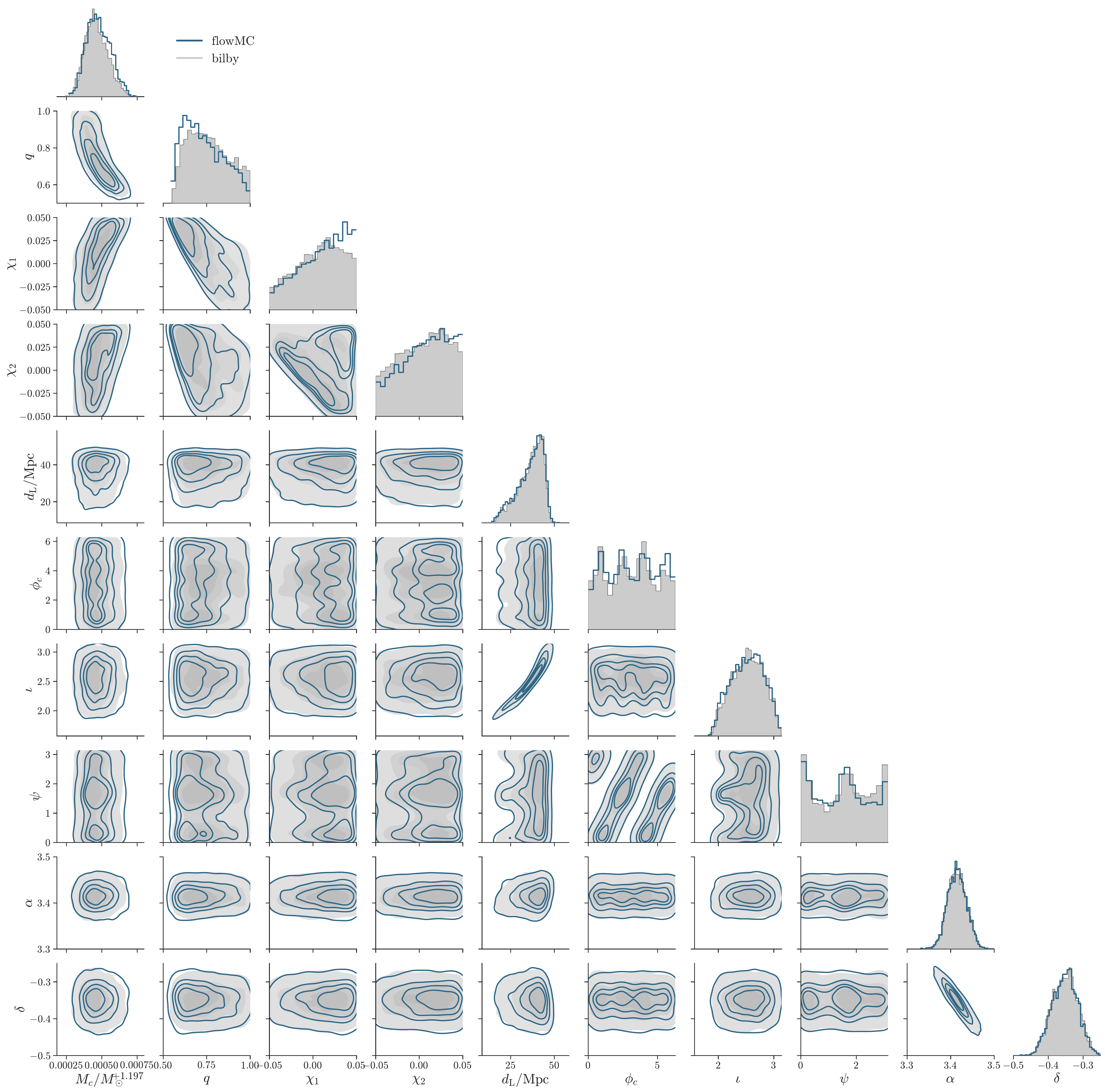}
\caption{
    GW170817 posterior computed by our code (blue) and \textsc{Bilby} (gray).
}
\label{fig:GW170817}
    \script{plot_GW170817.py}
\end{figure*}

For comparison, we produce equivalent runs with \textsc{Bilby}, using the same
exact data and priors.  We use the \textsc{dynesty} sampler
\citep{2020MNRAS.493.3132S,dynesty}, with 1000 live points and other settings as
in the configurations files available on github \url{https://github.com/kazewong/TurboPE/tree/main/src}.  We
carry out these runs using \textsc{parallel Bilby} (\textsc{pBilby})
\citep{Smith:2019ucc} to distribute the computation over 400 Intel Skylake CPUs
for each event.  For the specific settings chosen, the wall-time duration of
each run was ${\sim}2$ h for GW150914 and ${\sim}1$ day for GW170817.


Figs.~\ref{fig:GW150914} and \ref{fig:GW170817} show that our posteriors are
consistent with those produced by \textsc{Bilby}.  For a quantitative
comparison, we compute the Jensen-Shannon divergence between our code and
\textsc{Bilby} for the marginalized distribution for each parameter. The
Jensen-Shannon divergence (JSD) is a symmetric measure of the distance between
two probability distributions, with a value of 0 indicating identical
distributions and a value of $\ln{2}$ nat representing the maximum
possible divergence between two distributions. The JSD values for the two events
are shown in Table \ref{tab:JSD}. The maximum JSDs for GW150914 and GW170817 are
\jsdMaxBBH and \jsdMaxBNS, and the mean JSDs are \jsdAvgBBH and \jsdAvgBNS,
respectively. The JSD values are comparable to those reported in
\cite{Romero-Shaw:2020owr}, which show our code agrees with existing tools.

\begin{table}[hbt!]
    \begin{center}
  \input{output/js_table.tex}\label{output/js_table.tex}\unskip%

    \caption{JSD values for the marginalized distribution of each parameter for
    GW150914 and GW170817 between our code and \textsc{Bilby}. The bold values
    indicate the parameters with the largest JSD.}
    \label{tab:JSD}
    \end{center}
\end{table}

\section{Discussion}
\label{sec: Discussion}

\subsection{Comparison to other approaches}

There have been several recent efforts to speed up GW parameter estimation,
relying on techniques ranging from efficient reparameterizations
\citep{Islam:2022afg,Roulet:2022kot} to deep learning
\citep{Dax:2021tsq,Dax:2022pxd}. While all of these methods can achieve
minutes-scale parameter estimation with high fidelity under some conditions,
our approach possesses unique strengths, and may complement some of those other
techniques. Additional approaches for speeding up GW parameter estimation include
\cite{Canizares:2014fya,Smith:2016qas,Lee:2022jpn,Wofford:2022ykb,Lange:2018pyp,Williams:2021qyt,Morisaki:2021ngj};
here we discuss those most relevant in the context of our work.

In contrast to \cite{Dax:2021tsq,Dax:2022pxd}, we do not require pre-training
the neural network on a large collection of waveforms and noise realizations.
This means that our algorithm can be immediately deployed as soon as new
waveform models and noise models are available. Furthermore, our method is at
its core an MCMC algorithm, meaning it inherits the merit of convergence
measures in MCMC. As we are only using the normalizing flow as a proposal
distribution, and the normalizing flow is trained jointly with a local sampler,
we do not risk overfitting since our training data is being generated on the
fly and is always approaching the target distribution. In this sense, we do not
introduce potential extra systematic errors to the inference results.

While our pipeline uses samples generated by the local sampler for training,
one could also supply a pre-trained normalizing flow to our pipeline to bypass
the training stage. This would have the advantage of further reducing the total
runtime; however, it could introduce systematic bias in the inference result if
the pre-trained network is not able to capture the complexity presented in the
data.

In contrast to \cite{Islam:2022afg,Roulet:2022kot} or other reparameterization
schemes, we do not rely on handcrafted coordinate systems for sampling. If a
useful coordinate transformation is known ahead of time, it can be trivially
implemented within our pipeline, potentially easing convergence. However,
tailored reparameterizations rely on specific assumptions about the targeted
signal, which cannot always be generalized beyond specific applications. On the
other hand, within our pipeline, the normalizing flow effectively discovers
reparameterizations that ease sampling automatically without \emph{a priori}
knowledge of the structure of the problem. In general, the transformation
discovered by the normalizing flow will only be approximate and hence not as
efficient as an explicit reparameterization of the problem; yet, our approach
applies to a much border class of problems where clever coordinate
transformations are not known ahead of time, such as parameter estimation with
precessing waveforms, calibration parameters, testing GR, and multi-event joint
inference.

It is always beneficial to reparameterize if a convenient mapping is known
ahead of time. For the class of problems treated in
\cite{Islam:2022afg,Roulet:2022kot}, we can incorporate those
reparameterizations directly into our MCMC pipeline to reduce the complexity of
the problem, hence speeding up the training phase. If there are limitations to
the reparameterization that mean it cannot properly encompass part of the
target posterior, the normalizing flow should still be able to learn to produce
accurate samples efficiently.

The two avenues discussed here (machine learning and reparameterizations)
represent two orthogonal directions one can take in building next generation PE
tools. On the one hand, there are modern techniques such as deep learning that
are very flexible and powerful, but rely on having highly robust training data.
On the other hand, there are traditional tools that make use of our
understanding of the underlying physics to ease sampling, which relies
on having good intuition of the problem and tends to not generalize.
Our method achieves the key advantages of both approaches, without many of their limitations.

The main difference between methods used in industrial products and scientific
problems is that the latter must address questions that may not have been
answered before, hence requiring techniques that generalize beyond the current
state of knowledge robustly. Our work leverages both reparameterization and
machine learning, yet our method can be trivially extended to problems beyond
standard CBC analyses that would be unsuitable for reparameterizations or deep
learning alone. Beyond efficiency, such flexibility and robustness are crucial
for building scientific tools.

\subsection{Future development}

We are currently working on a number of improvements and extensions to our
current infrastructure. While the \textsc{IMRPhenomD} waveform approximant is a
reasonable start, it lacks some qualitative features that other
state-of-the-art models have, such as precession, subdominants moments of the
radiation, tides, or eccentricity. It also has a higher mismatch with reference
numerical relativity waveforms compared to more recent waveform models.
Currently, we are working on building differentiable implementations of
\textsc{IMRPhenomXPHM} \citep{Pratten:2020ceb}, a precessing successor to
\textsc{IMRPhenomD} including subdominant harmonics, as well as the numerical
relativity surrogate waveforms, including \textsc{NRSur7dq4}
\citep{Varma:2018mmi}. Going forward, we expect the use of autodifferentiation
environments like \textsc{Jax} to become more prevalent in the waveform
development community, increasing the number of differentiable waveform models
available. This would not only be beneficial for parameter estimation, but also
for a number of other applications such as Fisher matrix computations, template
placement and calibrating waveforms to numerical relativity \citep{Coogan:2022qxs,Iacovelli:2022bbs,Iacovelli:2022mbg,ripplepaper}.

While standard CBC analyses go up to 17 dimensions, non-standard GW PE problems
can have more parameters, which could potentially lead to more complicated
geometries in the target posterior that is hard to reparameterize. For example,
\cite{LIGOScientific:2021sio} introduces 10 extra parameters controlling
deviations in the post-Newtonian coefficients predicted in GR. On top of the
increase of dimensionality, these parameters often introduce non-trivial
degeneracies such as local correlation and multi-modality. Therefore, currently
testing GR is limited in practice to varying these modifications one at a time,
partially due to the bottleneck in the sampler.  Given the gradient-based and
normalizing flow-enhanced sampler, our code shows promise in tackling this
problem.

Our current code can perform parameter estimation for any combination of
ground-based detectors, under the assumption that signals are
transient and their wavelength is short. The first condition guarantees that
the effect of Earth's rotation can be ignored when computing antenna patterns,
while the second means that we can treat the antenna patterns as frequency
independent constants. These assumptions break for next-generation detectors,
whether on Earth or in space, like Cosmic Explorer, Einstein Telescope and
LISA; differentiable implementations of antenna patterns for those detectors is
work in progress.

Furthermore, our current implementation is minimal and we do not make use of
most standard ``tricks'' to accelerate sampling.  In particular, we do not
incorporate (semi)analytic marginalization schemes over parameters such as time
and phase \citep{2019PASA...36...10T}. As the performance of our implementation is not
significantly impacted by having two extra dimensions, time and phase
marginalization are not crucial for us;
however, their implementation within our framework would be trivial.

For simplicity, we did not include calibration uncertainties in this study, but
implementing this is also straightforward. Additionally,
we took the noise PSD as an input for our analysis, as is the case for most
traditional PE pipelines.  However, since we are planning to deploy this
pipeline to perform nearly-online PE, we also need to consider estimating the
PSD on the same timescale, for which \cite{Cornish:2021wxy} has proposed a
solution. We are looking into incorporating this, as well as analytic
marginalization schemes and calibration uncertainties, into our pipeline.

When it comes to wall time, the just-in-time compilation of our code is the
current limiting factor.  While \textsc{Jax}'s JIT compilation drastically
accelerates likelihood evaluations, it comes with a significant compilation
overhead before the first evaluation. We observe that the compilation time
depends on the device where the code is run; this is expected since
\textsc{Jax} leverages the \textsc{Accelerated Linear Algebra} (XLA) compiler
to take advantage of hardware accelerators, which means that \textsc{Jax} needs
to compile the code for each specific device according to its architecture. On
an \textsc{Nvidia} A100 GPU, the compilation overhead could go up to 3 minutes
for our current waveform. Meanwhile, for the cases we have studied, the time
needed to obtain converging sampling on an A100 is about $\sim 40\,$s. This
means the compilation overhead dominates the wall-clock time of our current PE
runs. To maximize the potential of our code, we are looking into ways to reduce
the compilation overhead or to cache the compilation results to avoid paying
the compilation overhead for every event.

Besides compilation, there is in principle also overhead from finding the
reference waveform used in for heterodyning the likelihood. Since the
\textsc{differential evolution} algorithm we currently use has not been
implemented in \textsc{Jax}, and the \textsc{Jax} waveform we use is not
compatible with the parallelization scheme in the \textsc{scipy} library,
maximizing the likelihood currently takes us around 1 minute for GW170817. There
are two ways to reduce this time.

First, we can explore a different optimization strategy that
takes full advantage of the strengths of our pipeline, in particular the
differentiability of our likelihood and the possibility to evaluate many
waveforms in parallel with a GPU. Particle swarm \citep{7869491} and stochastic
gradient descent methods \citep{10.5555/304710.304720} are promising candidates
we are investigating.

Second, we may marginalize extrinsic parameters to reduce the dimensionality of
the optimization problem. Currently, we simultaneously maximize all 11 CBC
parameters in our problem numerically, which is unnecessary. There are
long-existing, efficient maximization schemes for extrinsic parameters, such as
the merger time and phase, which can find the corresponding maximum likelihood
waveform much more efficiently when compared to differential evolution. We
expect implementing these schemes will reduce the time needed to find the
reference waveform parameters by fixing the extrinsic parameters and by reducing
the dimensionality of the optimization problem. Furthermore, the search pipeline
provides a subset of the parameters such as the masses, which can be fixed during the
optimization to further reduce the dimensionality of the problem.

Finally, one important aspect of modern computing is scalability, meaning it is
generally favorable if one can simply devote more computing units to the same
problem in order to reduce the wall time. In our case, this means that we would like to
use more than one GPU for the same PE process. More GPUs can help in the
following ways: first, we can run more independent chains at
the same time, which can generate more samples faster; second, and more
importantly, as shown in this work and \cite{2022arXiv221106397W}, more independent
chains also help reduce the burn-in time. Parallelizing over the number
of chains dimension is trivial and does not require much change to the current
infrastructure. Additional GPUs can also help by enabling faster training of
larger flow models. While the training time is not the biggest bottleneck given
the flow model used in this study, more GPUs means we can increase the
bandwidth of the flow model by increasing its size while keeping the training
time the same. This would help capture more complex geometries in the target
space, which can lead to better convergence in general.

\section{Conclusion}

In this work, we presented a PE pipeline for GW events that is efficient,
flexible and reliable. Our package brings together a number of innovations,
including differentiable waveform models, likelihood heterodyning, and
normalizing-flow enhanced gradient-based sampling. We tested the robustness of
our pipeline, currently built upon \textsc{ripple} and \textsc{flowMC}, on a
set of 1200 synthetic GW events, showing it is robust, unbiased and efficient
enough to handle the large catalogs of detections that will be available in the
near future. We also show that our pipeline can estimate the parameters of
GW150914 and GW170817 within a couple minutes, demonstrating the potential of our
implementation on real data.

\section{Acknowledgements}
We thank Will M.~Farr, Aaron Zimmerman, Daniel Foreman-Mackey and Marylou Gabri\'e for helpful discussions; we also thank Carl-Johan Haster, Neil J.~Cornish and Thomas Dent for comments on the draft.
The Flatiron Institute is a division of the Simons Foundation.
This material is based upon work supported by NSF's LIGO Laboratory which is a major facility fully funded by the National Science Foundation.
This research has made use of data or software obtained from the Gravitational Wave Open Science Center (gw-openscience.org), a service of LIGO Laboratory, the LIGO Scientific Collaboration, the Virgo Collaboration, and KAGRA. LIGO Laboratory and Advanced LIGO are funded by the United States National Science Foundation (NSF) as well as the Science and Technology Facilities Council (STFC) of the United Kingdom, the Max-Planck-Society (MPS), and the State of Niedersachsen/Germany for support of the construction of Advanced LIGO and construction and operation of the GEO600 detector. Additional support for Advanced LIGO was provided by the Australian Research Council. Virgo is funded, through the European Gravitational Observatory (EGO), by the French Centre National de Recherche Scientifique (CNRS), the Italian Istituto Nazionale di Fisica Nucleare (INFN) and the Dutch Nikhef, with contributions by institutions from Belgium, Germany, Greece, Hungary, Ireland, Japan, Monaco, Poland, Portugal, Spain. The construction and operation of KAGRA are funded by Ministry of Education, Culture, Sports, Science and Technology (MEXT), and Japan Society for the Promotion of Science (JSPS), National Research Foundation (NRF) and Ministry of Science and ICT (MSIT) in Korea, Academia Sinica (AS) and the Ministry of Science and Technology (MoST) in Taiwan.
T.E.\ is supported by the Horizon Postdoctoral Fellowship.

\bibliography{bib}

\end{document}

%% file: output/macros.tex
\renewcommand{\ptot}{0.70\xspace}
\renewcommand{\jsdMaxBBH}{0.0172\xspace}
\renewcommand{\jsdAvgBBH}{0.0031\xspace}
\renewcommand{\jsdMaxBNS}{0.0073\xspace}
\renewcommand{\jsdAvgBNS}{0.0024\xspace}

%% file: output/js_table.tex
\begin{tabular}{lrr}
\toprule
 & GW150914 & GW170817 \\
\midrule
$M_c$ & \textbf{0.01716} & 0.00418 \\
$q$ & 0.00361 & \textbf{0.00732} \\
$\chi_1$ & 0.00234 & 0.00354 \\
$\chi_2$ & 0.00099 & 0.00179 \\
$d_{\rm{L}}$ & 0.00095 & 0.00055 \\
$\phi_c$ & 0.00032 & 0.00195 \\
$\iota$ & 0.00262 & 0.00128 \\
$\psi$ & 0.00056 & 0.00175 \\
$\alpha$ & 0.00073 & 0.00102 \\
$\delta$ & 0.00202 & 0.00111 \\
\bottomrule
\end{tabular}